\documentclass[conference]{IEEEtran}

\usepackage{cite} 
\usepackage{amsmath,amssymb,amsfonts}
\usepackage{adjustbox} 
\usepackage[ruled,vlined,linesnumbered]{algorithm2e}
\usepackage{graphicx}
\usepackage{booktabs}
\usepackage{multirow}
\usepackage{caption}
\usepackage{float}
\usepackage{tikz}
\usetikzlibrary{arrows.meta,positioning,shapes.geometric}
\usepackage{url}
\usepackage{mdframed}
\usepackage{tabularx}
\usepackage{makecell}

\title{Seeing Radio: From Zero RF Priors to Explainable Modulation Recognition with Vision Language Models}

\author{%
\IEEEauthorblockN{%
Hang~Zou\IEEEauthorrefmark{1},
Bohao~Wang\IEEEauthorrefmark{1}\IEEEauthorrefmark{2},
Yu~Tian\IEEEauthorrefmark{1},
Lina~Bariah\IEEEauthorrefmark{1},
Chongwen~Huang\IEEEauthorrefmark{2}, 
Samson~Lasaulce\IEEEauthorrefmark{3},
Mérouane~Debbah\IEEEauthorrefmark{1}}
\IEEEauthorblockA{\IEEEauthorrefmark{1} Research Institute for Digital Future, Khalifa University, 127788 Abu Dhabi, UAE}
\IEEEauthorblockA{\IEEEauthorrefmark{2} College of Information Science and Electronic Engineering, Zhejiang University, 310027, Hangzhou, China}
\IEEEauthorblockA{\IEEEauthorrefmark{3}Universit\'{e} de Lorraine, CNRS, CRAN, F-54000 Nancy, France}
}

\begin{document}
\addtolength{\topmargin}{0.03in}
\addtolength{\textheight}{-0.03in}

\maketitle

\begin{abstract}
Current RF machine-learning pipelines rely on task-specific deep networks for modulation classification and related tasks, but these models require custom architectures and labeled datasets for each problem, generalize poorly across channel conditions and SNRs, and offer little interpretability. In contrast, modern multimodal large language models (MLLMs) can integrate heterogeneous visual and textual data and exhibit strong cross-domain generalization and explanation capabilities. Our goal in this work is to explore whether vision–language models (VLMs) can be adapted to directly perceive RF signals and reason about modulation patterns without redesigning their architectures or injecting RF-specific inductive biases. To achieve this, we convert complex IQ streams into time-series, spectrogram, and joint RF visualizations, build a 57-class RF visual question answering benchmark, and show that lightweight parameter-efficient fine-tuning can enhance the accuracy of a general-purpose VLM from around 10\% to nearly 90\%, with ensuring robustness to noise and out-of-vocabulary modulations and the ability to produce human-readable rationales. The obtained results show that combining RF-to-image conversion with promptable VLMs provides a scalable and practical foundation for RF-aware AI systems in future 6G networks.
\end{abstract}



\section{Introduction}

Large language models (LLMs) have evolved rapidly in both scale and capability, becoming increasingly versatile across diverse applications. Through scaling laws and instruction tuning, they have enabled outstanding general-purpose reasoning and strong few-shot learning performance across a wide range of tasks~\cite{brown2020gpt3}. Building on this progress, multimodal LLMs have emerged, bridging vision and language by pairing frozen or jointly trained vision encoders with powerful LLM backbones to ground linguistic reasoning in visual context~\cite{liu2023llava,bai2025qwen25vl,zhu2025internvl3}. As this evolution accelerates, foundation models such as GPT-5 demonstrate robust multimodal comprehension and strong cross-domain generalization, integrating textual and visual knowledge to solve more complex reasoning tasks~\cite{openai20245gpt5}.

In a parallel race, the 6G research agenda anticipates AI-native networks that can orchestrate sensing, communications, computing, and control at unprecedented scales and latencies. Accordingly, existing surveys and position papers emphasize on the foundational integration of advanced AI technologies throughout the entire network stack, from radio access to core network automation, network slicing, and semantic communication~\cite{zhao2020comprehensive,6gIntelligentNetwork2024}. Within this vision, LLMs emerge as unified interfaces for knowledge access, tool-use, reasoning and planning, supporting automated policy optimization, intelligent fault diagnosis, and intent-driven network operations for complex, heterogeneous infrastructures.

The early research on \textit{LLM4Telecom} have explored LLM-driven network and service management assistants, domain-specialized instruction-tuned models, and agentic workflows that integrate LLMs into network tools and monitoring data~\cite{wirelessllm2024,bariah2024next,hao2025CST}. These studies have showcased promising results in network troubleshooting, configuration generation, root cause analysis, and closed-loop operations. However, they mainly rely on language, in the form of text, tables or structured logs, as the primary modality. Within this direction, Telecom-specific LLMs have been developed, such as TelecomGPT \cite{zou2025telecomgpt}, to improve LLMs capabilities in telecom related tasks. Instead of relying on general-purpose LLMs or prompt engineering, telecom knowledge has been directly embedded into the LLMs to equip them with the needed operational intelligence, domain-grounded reasoning, and contextual understanding that are essential to reason over network scenarios and automate management tasks.

However, despite all this progress in Telecom LLMs, currently their integration into telecom networks is largely text-centric. Hence, this prevents the exploitation of such powerful tools in more advanced network tasks that require that an LLM can process and generate different kind of data modalities, including radio frequency (RF) signals, images, or acoustic signals, among others. It is worthy to note that future networks will not be only conversation- and reasoning-dependent, but they will also enjoy advanced RF perception capabilities.

Existing machine learning-driven RF intelligence is dominated by narrow, task-specific models for tasks such as automatic modulation classification (AMC), channel estimation, beam selection, interference identification, and spectrum sensing. These models are typically trained on small, heterogeneous datasets under controlled assumptions about channels, hardware impairments, and traffic patterns. While such designs can achieve high accuracy within their training distributions, they often rely on separate task-specific architectures and datasets, require substantial expert-labeled data, generalize poorly across varying SNRs and deployment conditions, and provide limited interpretability or interaction beyond a single hard prediction.

At the same time, modern vision--language models (VLMs) have demonstrated remarkable cross-domain generalization and multimodal reasoning abilities, for example when operating on audio spectrograms~\cite{dixit2024vision}. Yet, to date, no VLM has been explicitly enabled with \emph{spectrum intelligence}, i.e., the ability to interpret or classify RF signal patterns that underpin key functions such as AMC, interference detection and spectrum sensing, all of which are crucial for dynamic spectrum access, cognitive radio, and autonomous network optimization. This gap motivates our central question: can we leverage off-the-shelf VLMs to acquire RF understanding and perception by working purely in the visual domain, without redesigning architectures or training from scratch?

\begin{figure}[t!]
\vspace{0.1in}
\centering
\includegraphics[width=0.9\linewidth]{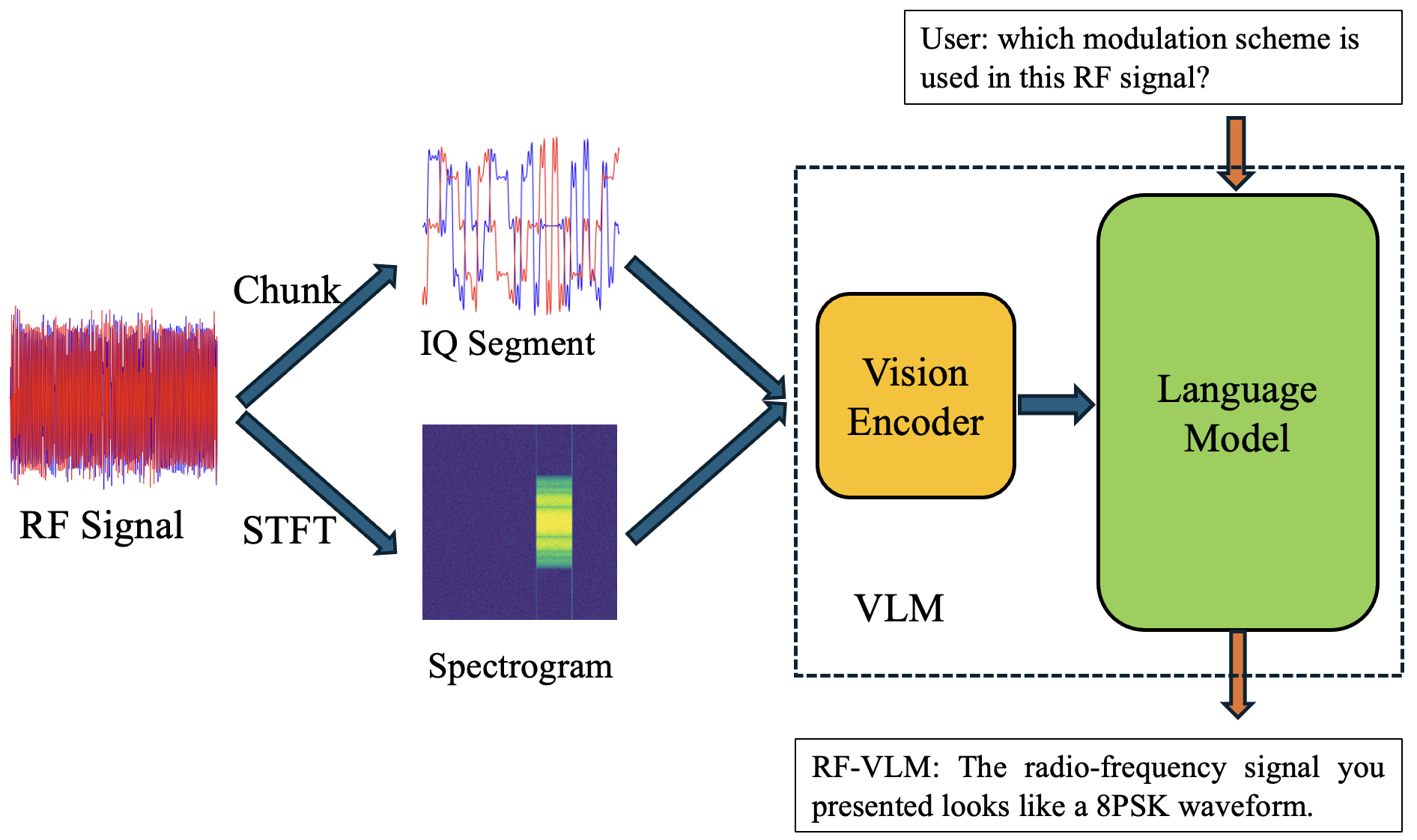}
\caption{Proposed framework. Spectrograms are obtained through STFT while IQ segment are obtained by down conversion and chunking. The VLM will be provided with image illustrating spectrogram, IQ segment, or both of them.}
\label{fig:method}
\end{figure}

\textbf{This paper proposes a simple, practical pipeline to enable off-the-shelf VLMs to understand raw RF IQ data, by converting them into visual contents that are interpretable by VLMs.} Specifically, we transform RF signals into spectrograms using short-time Fourier transform (STFT) and into and as chunked IQ time-series panels, to construct a visual-question-answering (VQA) dataset, comprising queries on exact modulation classes or higher-level signal categories. With lightweight fine-tuning and no architectural changes, we show that a modern VLM fine-tuned only on these images can perform \emph{accurate modulation recognition while providing natural language explanations of their decisions}. Our main contributions are summarized as follows:
\begin{itemize}
    \item We construct an efficient RF modulation classification benchmark covering spectrogram, time-series, and joint signal representations.
    \item We demonstrate, to the best of the authors' knowledge for the first time, that a general-purpose VLM can be adapted to classify RF modulations while producing concise, human-readable explanations.
    \item We show that, after minimal fine-tuning, the model achieves strong data efficiency and robustness to noise and out-of-vocabulary (OOV) modulation classes, paving the way for RF-aware multimodal LLMs for 6G.
\end{itemize}

\section{Benchmark Design for RF Modulation Classification}
\label{sec:rf-benchmark}

To assess the capability of VLMs in RF signal understanding, we construct a reproducible benchmark that evaluates VLMs on recognition of digital and analog modulation types from rendered RF visualizations. We leverage \emph{TorchSig} \cite{boegner2022torchsig} which provides 57 different modulation schemes across different families as follows: 

\begin{itemize}
\item \textbf{OFDM (12):} Orthogonal Frequency–Division Multiplexing with different number of subcarriers  \(M \in \{64,72,128,180,256,300,512,600,900,1024\}\) and \(  \{1200, 2048 \}\).

\item \textbf{QAM (8):} Quadrature Amplitude Modulation with constellation size $M \in \{16,32,64,256,1024\}$ plus cross-constellations: $\{32\_\text{cross},128\_\text{cross},512\_\text{cross}\}$.

\item \textbf{PSK (6):} Phase Shift Keying with different number of phase states $M \in \{2\,( \text{BPSK}),4\,(\text{QPSK}),8,16,32,64\}$.

\item \textbf{ASK (5):} Amplitude Shift Keying with different number of amplitude levels $M  \in \{4,8,16,32,64\}$.

\item \textbf{FSK (4):} Frequency Shift Keying with different number of distinct tones $M \in \{2,4,8,16\}$.

\item \textbf{GFSK (4):} Gaussian-filtered FSK with different number of distinct tones $M \in \{2,4,8,16\}$.

\item \textbf{MSK (4):} Minimum Shift Keying, constant-envelope CPM with different symbol alphabet size $M \in \{2,4,8,16\}$.

\item \textbf{GMSK (4):} Gaussian-filtered MSK  with different symbol alphabet size $M \in \{2,4,8,16\}$.

\item \textbf{AM (4):} Amplitude Modulation (Analog) with different sideband type in Double-Sideband with carrier, Double-Sideband, Suppressed Carrier and Single-Sideband AM (Upper / Lower sideband only).

\item \textbf{Chirp/LFM (3):} Linear Frequency Modulation with data-dependent chirp, Linear FM radar-style pulses and Chirp Spread Spectrum.

\item \textbf{Singleton (3):} Frequency Modulation (FM), On–Off Keying (OOK) and  un-modulated tone (Tone).

\end{itemize}

The benchmark supports three input image modes, namely, \emph{IQ segment only (\textbf{IQ})}, \emph{spectrogram only (\textbf{Spec})}, and \emph{joint spectrogram and IQ (\textbf{Joint})}, with deterministic pipelines for data synthesis, preprocessing and prompt design.

\subsection{RF preprocessors and image input modes}
\label{subsec:modalities}

\textbf{IQ segment classification} Given a complex baseband sequence of length $N$, $x[n] \in \mathbb{C}$, which can be obtained by down conversion of an RF signal, we visualize a short segment of the \emph{in-phase} ($\Re\{x[n]\}$) and \emph{quadrature} ($\Im\{x[n]\}$) components as two time series traces. To extract the underlying patterns from the RF waveform, we define zero-crossing points with a predefined hysteresis $\epsilon$:
\begin{equation}
\mathrm{ZC} \;=\; \{\, n \mid \Re\{x[n-1]\} \le \epsilon,~ \Re\{x[n]\} > \epsilon \,\}.
\end{equation}

\vspace{0.1in}
These zero-crossing points capture local oscillatory behavior and phase transitions, providing patterns that can be visually represented and classified by a VLM. Given integers $N_{\min}$ and $N_{\max}$, we sample an integer
$P \sim \mathcal{U}\{N_{\min},\,N_{\max}\}$ and extract the shortest
contiguous segment that contains
$P$ zero–crossing intervals. We avoid a fixed sample length because different carrier/symbol rates yield different oscillation rates, hence, a fixed window will show a variable number of cycles and distort the waveform’s visual geometry across modulation classes.
By counting periods instead, we normalize the time axis and ensure that each segment contains a consistent number of signal cycles, regardless of symbol rate or sampling configuration. Note that this preprocessing only selects the segment, i.e., the VLM still receives the raw IQ traces for that window (no additional temporal filtering or phase decoding).

\textbf{Spectrogram classification} We compute an STFT with window $w[n]$, fast Fourier transform (FFT) size $K$, and hop $H$,
\begin{equation}
S[k,t] \;=\; \sum_{n} x[n]\, w[n-tH]\, e^{-j2\pi kn/K},\ k\in[0,K{-}1]
\end{equation}
A Blackman window is used, with no centering and an FFT shift applied along the frequency axis. We render a magnitude spectrogram \( |S[k,t]|\) in dB, followed by a perceptually uniform colormap (e.g., \texttt{viridis}), ensuring that amplitude variations are visually consistent and comparable across modulation types. This time–frequency representation gives unique spectral patterns, such as symbol transitions, bandwidth occupancy, and sideband structures, allowing the VLM to differentiate between modulation schemes through visual hints rather than explicit signal models.

\textbf{Joint spectrogram and IQ segment classification} We simply concatenate the spectrogram with an IQ segment along the width. Each image consists of two horizontal panels, namely magnitude spectrogram on left and IQ time-series on right (real in blue, imag in red). The panels are height-matched before concatenation.

The resolutions of all images are aligned with the FFT size to avoid compression error introduced during resizing. For both spectrograms and IQ segments no time or frequency axis are provided so that the VLM will focus on the pattern itself and avoid leaking from legend for all image modes. The proposed method is summarized in Fig. \ref{fig:method}. 

\subsection{Prompt Design}
\label{subsec:episodes}

We support both \emph{zero-shot} (with no examples provided) and \emph{few-shot} (with few examples provided) evaluations with $n$-way (selecting the correct one among $n$ choices) classification:
\begin{itemize}
\item \textbf{Zero-shot $n$-way:} Each record contains (i) a \emph{query} image, (ii) a list of $n$ candidate classes, uniformly sampled without replacement (balanced target cycling), and (iii) the gold label.
\item \textbf{Few-shot $n$-way, $s$-shots:} For each candidate class we attach $s$ example images (disjoint from the query) before the query image. Examples and queries are drawn from different splits to avoid data leakage.
\end{itemize}

\begin{figure}[t!]
\vspace{0.05in}
\centering
\begin{mdframed}[linecolor=black, linewidth=2pt, roundcorner=10pt]
\textbf{VQA Prompt Template for Modulation Classification}
\hrule height 0.5pt
\vspace{0.15in}

\textbf{System:} You are a helpful assistant with expertise in recognizing patterns and identifying RF modulations from visual inputs. 

\vspace{0.08in}
\textbf{User:} Your task is to analyze an RF visualization and determine the most likely modulation class from a given list.  Note: this image has two horizontal panels. LEFT: magnitude spectrogram (dB), RIGHT: a short IQ time-series panel showing \emph{real} part (\textcolor{blue}{blue}) and \emph{imaginary} part (\textcolor{red}{red}) of the signal plotted over time. The time-series corresponds to a segment down converted from the RF waveform. Here are the classes: \textbf{[‘am-dsb’, ‘ofdm-128’, ‘16gmsk’, ‘4fsk’, ‘ook’, ‘16ask’, ‘16psk’, ‘8ask’, ‘32ask’, ‘ofdm-300’]}.  Your response must contain the exact name of the class only. Here is the \textbf{{IMAGE}}:

\vspace{0.08in}
\begin{center}
  \includegraphics[width=0.7\linewidth]{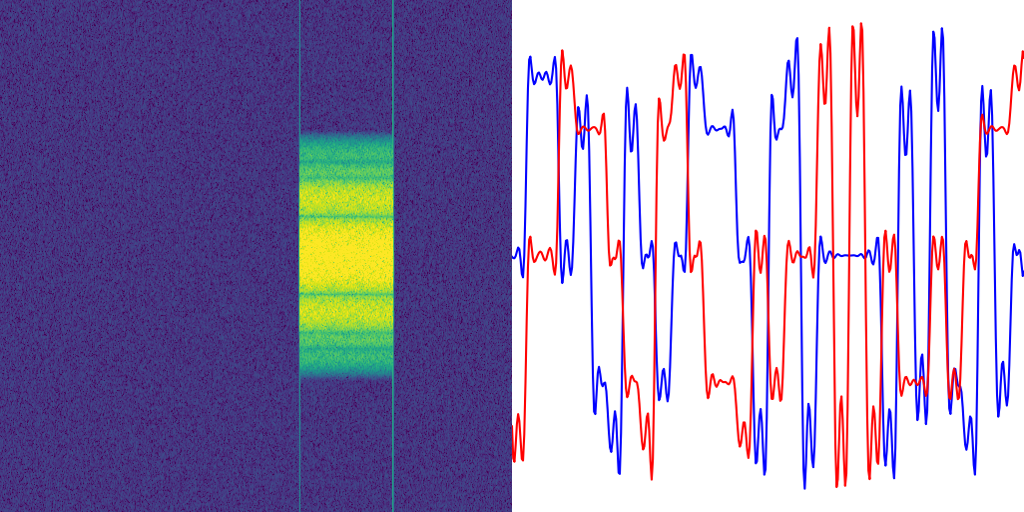}
\end{center}

\vspace{0.08in}
\textbf{VLM:} $\left\{ \text{modulation-class} \right\}$
\end{mdframed}
\caption{VQA-style prompt template for zero-shot RF modulation \(10\)-way classification provided a concatenation of spectrogram magnitude and the IQ time series segment.}
\label{fig:mcq_vqa_template}
\end{figure}

In practical wireless systems, modulation schemes are selected from a finite, predefined set of options, that are defined by the standards, frequency bands, or services. Hence, the set of admissible schemes (MCSs) is small and fixed, and additional context (e.g., LTE vs. Wi‑Fi, uplink) further narrows the options. Our $n$-way VQA therefore reflects practical settings, asking the model to pick the right modulation from a realistic shortlist rather than from an open‑ended label space. We use a simple chat-style schema with \texttt{system} and \texttt{user} roles to construct inputs for VLMs:

\textbf{Zero-shot prompts} The \texttt{system} prompt states that the model will see RF signals (spectrograms and/or IQ panels) and must classify from a given list. The \texttt{user} message contains:
\begin{enumerate}
\item A concise instruction: \emph{``Your task is to analyze an RF visualization and determine the most likely modulation class in the following list: [C\textsubscript{1}, \dots, C\textsubscript{n}].''}
\item A \emph{mode-specific note}, automatically attached based on the image mode:
    \begin{itemize}
    \item \textbf{Spec}: includes only the spectrogram magnitude, without phase info or no time-domain waveform.
    \item \textbf{IQ}: includes a short IQ time-series panel showing real part (\textcolor{blue}{blue}) and imaginary part (\textcolor{red}{red}) of the signal plotted over time,  without spectrogram views.
    \item \textbf{Joint}: includes a combination of both representations; encouraging the model to combine spectral and temporal patterns.
    \end{itemize}
\end{enumerate}

\textbf{Few-shot prompts:} To help the model ground its reasoning, we begin with a short introductory instruction: \emph{``You will see examples for several classes, followed by a query image. Use frequency patterns (spectrogram) and temporal dynamics (IQ waveforms) as appropriate.''} Then, for each candidate class $c_i$, we add $s$ example images labeled \emph{``Spectrogram (and/or time-series panels) for $c_i$"}. Finally, we add the query block with the same instruction as zero-shot and the query image(s). VLMs are always required to return only the exact class name.

\section{VLMs Are RF Signal Classifiers After Fine‑Tuning}
\label{sec:method}
\subsection{General-purpose VLMs Fail to Generalize to RF Domains}
We start with benchmarks built with \emph{noiseless} RF signals. We set FFT size \(K =512\) with hop \(H = 256\), with each time series waveform contains 20-25 zero-crossing intervals. We evaluate popular commercial and open-sourced VLMs on our benchmark over \(10\)-way \( 1000\) VQAs for each image mode. For commercial VLMs, we consider GPT5-nano \cite{openai20245gpt5} while Qwen2.5-VL-7B-Instruct \cite{bai2025qwen25vl} and InternVL3-8B \cite{zhu2025internvl3} are selected among open-sourced ones. We report the zero-shot and few-shots evaluations of the above models in Table \ref{tab:rf-vlm-shots}. We first observe that considered VLMs possess almost no prior knowledge of the spectral or temporal representations of RF signals and struggle to learn such domain-specific patterns through few-shot in-context learning. These results indicate that pretrained VLMs lack usable priors for RF time–frequency patterns and do not acquire them via in-context prompting alone. Next, we examine whether lightweight, parameter-efficient fine-tuning (PEFT) can close this gap.

\begin{table}[t]
\vspace{0.12in}
\centering
\resizebox{\linewidth}{!}{%
\begin{tabular}{l l c c c}
\toprule
\textbf{Model} & \textbf{Image Mode} & \textbf{Zero-shot} & \textbf{1-shot} & \textbf{2-shot} \\
\midrule
\multirow{3}{*}{InternVL3-8B}
 & \makecell{Spec}    &  9.70\% & 20.75\% & 24.67\% \\
 & \makecell{IQ}    & 10.70\% & 23.68\% & 26.84\% \\
 & \makecell{Joint}    & 9.80\% & 21.45\% & 26.95\%    \\
\midrule
\multirow{3}{*}{\makecell[l]{\footnotesize Qwen2.5-VL-7B-\\Instruct}}
 & \makecell{Spec}   & 12.68\% & 24.76\% & 30.71\% \\
 & \makecell{IQ}    & 15.40\% & 32.33\% & 36.25\% \\
 & \makecell{Joint}  & 11.50\% & 27.65\% & 29.45\%    \\
\midrule
\multirow{3}{*}{GPT5-Nano}
 & \makecell{Spec}   & 18.2\%  & 30.0\%  & 27.8\%  \\
 & \makecell{IQ}    & 17.1\%  & 18.1\%  & 21.2\%  \\
 & \makecell{Joint}  & 16.0\%  & 31.4\%  & 22.2 \%    \\
\bottomrule
\end{tabular}%
}
\caption{Overall accuracy (\%) of VLMs on RF modulation classification across different shots and image modes.}
\label{tab:rf-vlm-shots}
\end{table}

\subsection{PEFT Enables RF Signal Recognition}

To quantify domain adaptation ability of VLMs, we apply PEFT to VLMs using QLoRA~\cite{dettmers2023qlora}, which has proven effective for adapting LLMs to specialized domains. We fine-tune two open-sourced VLMs using a workstation equipped with two Nvidia A6000 GPUs. The LoRA rank and scaling factor are fixed at \(r=32\) and \(\alpha=64\), respectively. Each model is trained for 5 epochs with a learning rate of \(2\times10^{-4}\) on a dataset containing all image modes and 57 modulation classes, totaling 40K samples, approximately 700 samples per class. We denote the fine-tuned VLMs with the suffix ``-RF".

As shown in Table~\ref{tab:perfamlily}, fine-tuning leads to substantial accuracy gains across all visual representations from around 10\% to around 80\%. For Qwen2.5-VL-7B-Instruct-RF, the joint representation achieves 89.05\% accuracy, outperforming both the spectrogram-only (77.70\%) and IQ-only representations (77.20\%). Similarly, InternVL3-8B-RF reaches the highest 91.20\% with joint representation. Across all modulation families and visual representations, the joint representation consistently outperforms individual modalities. In particular, it significantly boosts weaker IQ-only families, such as OFDM (from 43.21\% to 88.43\%), and AM (from 50.00\% to 95.06\%), while maintaining strong performance on spectrally distinctive families, including FSK, GFSK, and Chirp/LFM. These results demonstrate that fusing spectral and temporal information allows VLMs to jointly capture frequency-domain stability and time-domain dynamics, leading to more discriminative and generalizable modulation features. The joint representation therefore provides a strong basis for extending the proposed framework toward modality-robust and domain-adaptive understanding of RF signals for VLMs.

\begin{table*}[t!]
\vspace{0.10in}
\centering
\setlength{\tabcolsep}{4.5pt}        
\renewcommand{\arraystretch}{1.12}    
\begin{adjustbox}{max width=\textwidth}
\begin{tabular}{llcccccccccccccc}
\toprule
 \makecell{\textbf{Model}} & \textbf{\makecell{Image \\ Mode}} & \textbf{w/o FT} & \textbf{With FT} & \textbf{ASK} & \textbf{FSK} & \textbf{GFSK} & \textbf{GMSK} & \textbf{MSK} & \textbf{QAM} & \textbf{PSK} & \textbf{OFDM} & \textbf{AM} & \textbf{Chirp/LFM} & \textbf{OOK} & \textbf{Tone} \\
\midrule
\multirow{3}{*}{\makecell{Qwen2.5-VL-7B\\Instruct}}
& Spec    & 12.68\% & 77.70\% & 71.43\% & 95.72\% & 90.00\% & \textbf{90.00\%} & 97.88\% & 46.79\% & 33.04\% & \textbf{88.67\%} & 91.49\% & 94.29\% & 97.14\% & 100.00\% \\
& IQ      & 15.40\% & 77.20\% & 79.43\% & 97.86\% & 90.71\% & 87.86\% & 95.71\% & \textbf{80.71\%} & \textbf{95.30\%} & 43.21\% & 50.00\% & 94.30\% & 100.00\% & 100.00\% \\
& Joint   & 11.50\% & \textbf{89.50\%} & \textbf{84.00\%} & \textbf{98.57\%} & \textbf{95.71\%} & 87.14\% & \textbf{98.57\%} & 71.43\% & 92.89\% & 88.43\% & \textbf{95.06\%} & \textbf{97.86\%} & \textbf{100.00\%} & \textbf{100.00\%} \\
\midrule
\multirow{3}{*}{\makecell{InternVL3-8B}}
& Spec    & 9.70\% & 77.35\% & 66.29\% & \textbf{98.57\%} & 94.29\% & \textbf{91.43\%} & \textbf{100.00\%} & 45.61\% & 35.61\% & 86.45\% & 88.57\% & 92.88\% & 100.00\% & 100.00\% \\
& IQ      & 10.70\% & 77.60\% & 78.86\% & 94.29\% & 89.29\% & 88.57\% & 98.57\% & 82.97\% & 86.37\% & 45.12\% & 72.86\% & 83.69\% & 97.14\% & 100.00\% \\
& Joint   & 9.80\% & \textbf{91.20\%} & \textbf{82.28\%} & 97.86\% & \textbf{94.29\%} & 90.71\% & 99.29\% & \textbf{85.06\%} & \textbf{93.45\%} & \textbf{90.24\%} & \textbf{92.14\%} & \textbf{95.04\%} & \textbf{100.00\%} & \textbf{100.00\%} \\
\bottomrule
\end{tabular}
\end{adjustbox}
\caption{Average accuracy (\%) with and without finetuning (FT) and per-family accuracies (\%) of VLMs.}
\label{tab:perfamlily}
\end{table*}


\subsection{Noise Robustness Evaluation}
\begin{figure}[t!]
    \vspace{0.12in}
    \centering
    \includegraphics[width=0.8\linewidth]{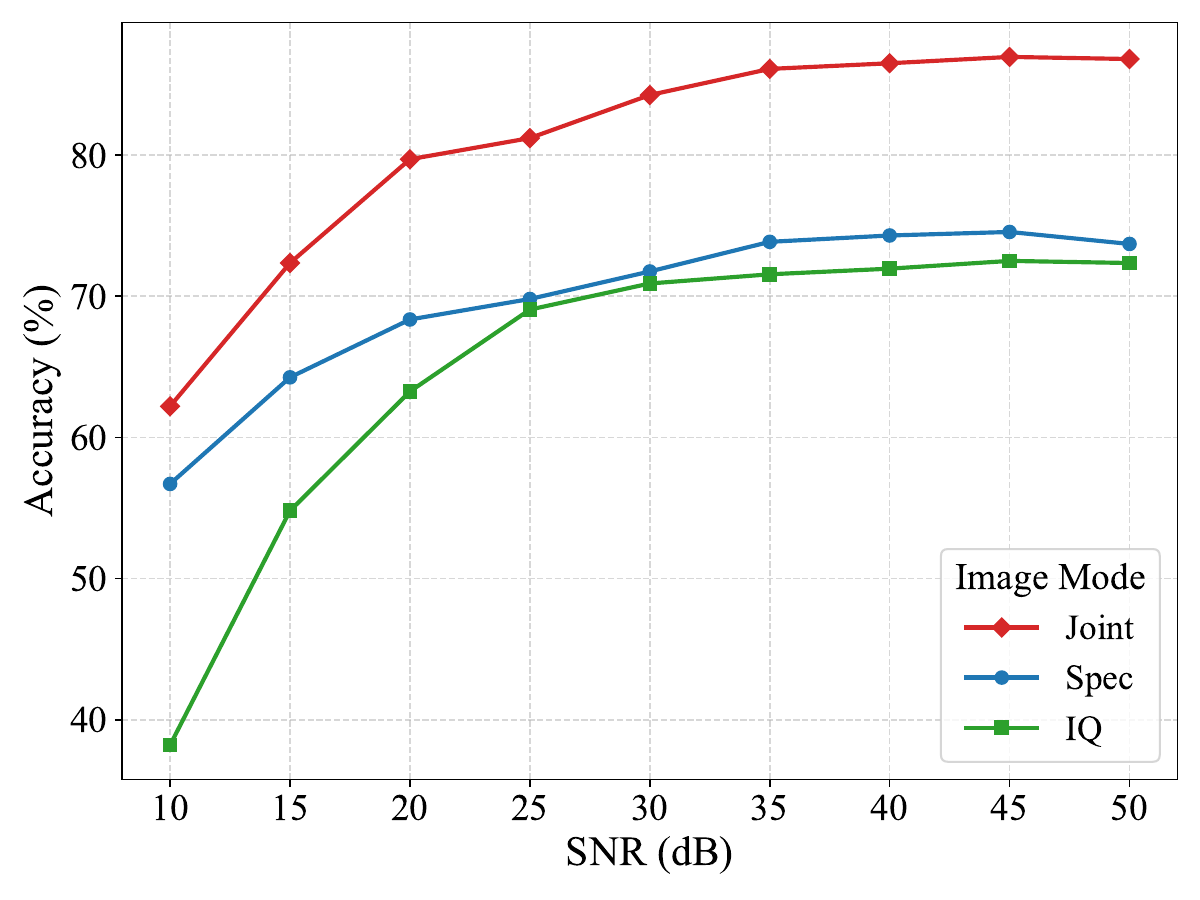}
    \caption{Accuracy vs. SNR of Qwen2.5-VL-7B-Instruct-RF under different image modes. }
    \label{fig:SNR}
\end{figure}


\begin{figure}[t!]
    \centering
    \includegraphics[width=0.8\linewidth]{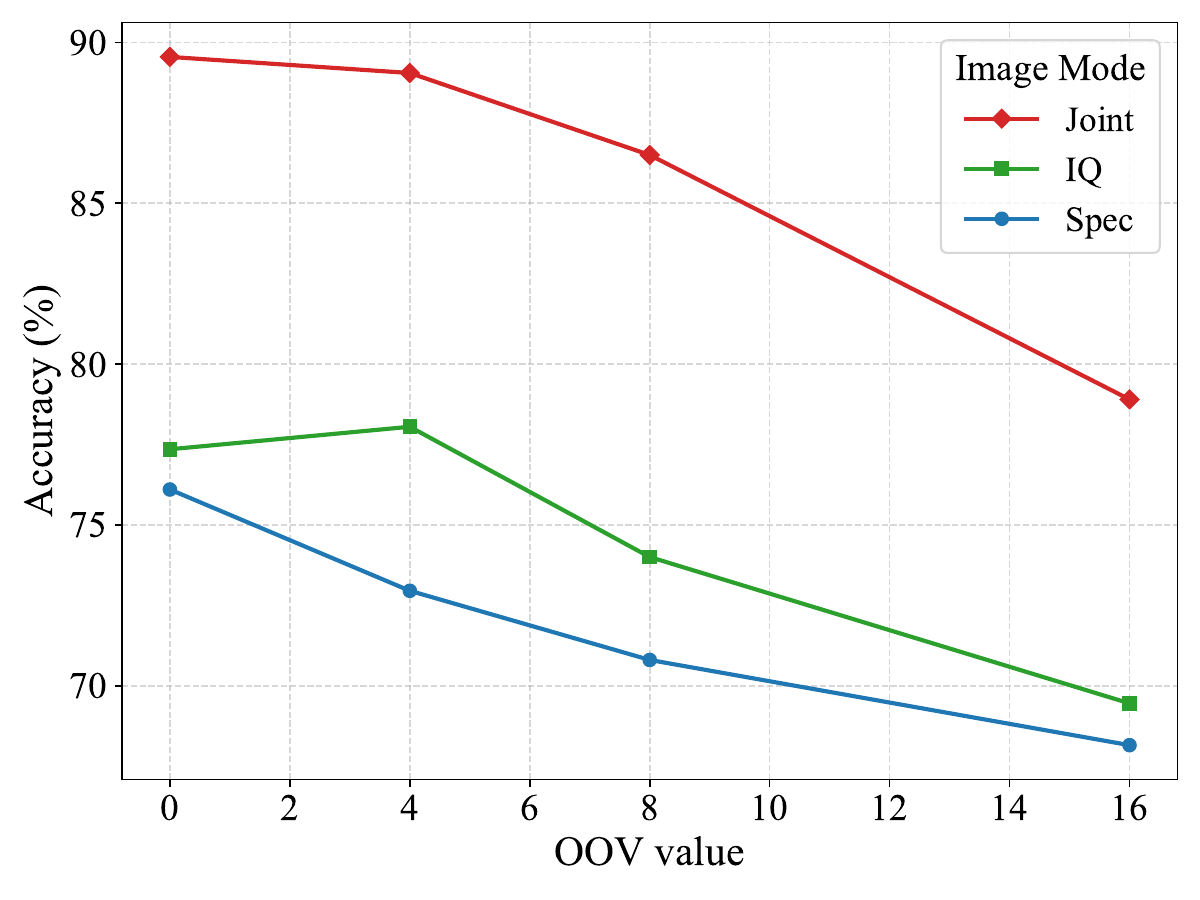}
    \caption{Accuracy vs. number of OOV classes of Qwen2.5-VL-7B-Instruct-RF under different image modes.}
    \label{fig:OOV}
\end{figure}

\begin{figure}[t!]
    \centering
    \includegraphics[width=0.8\linewidth]{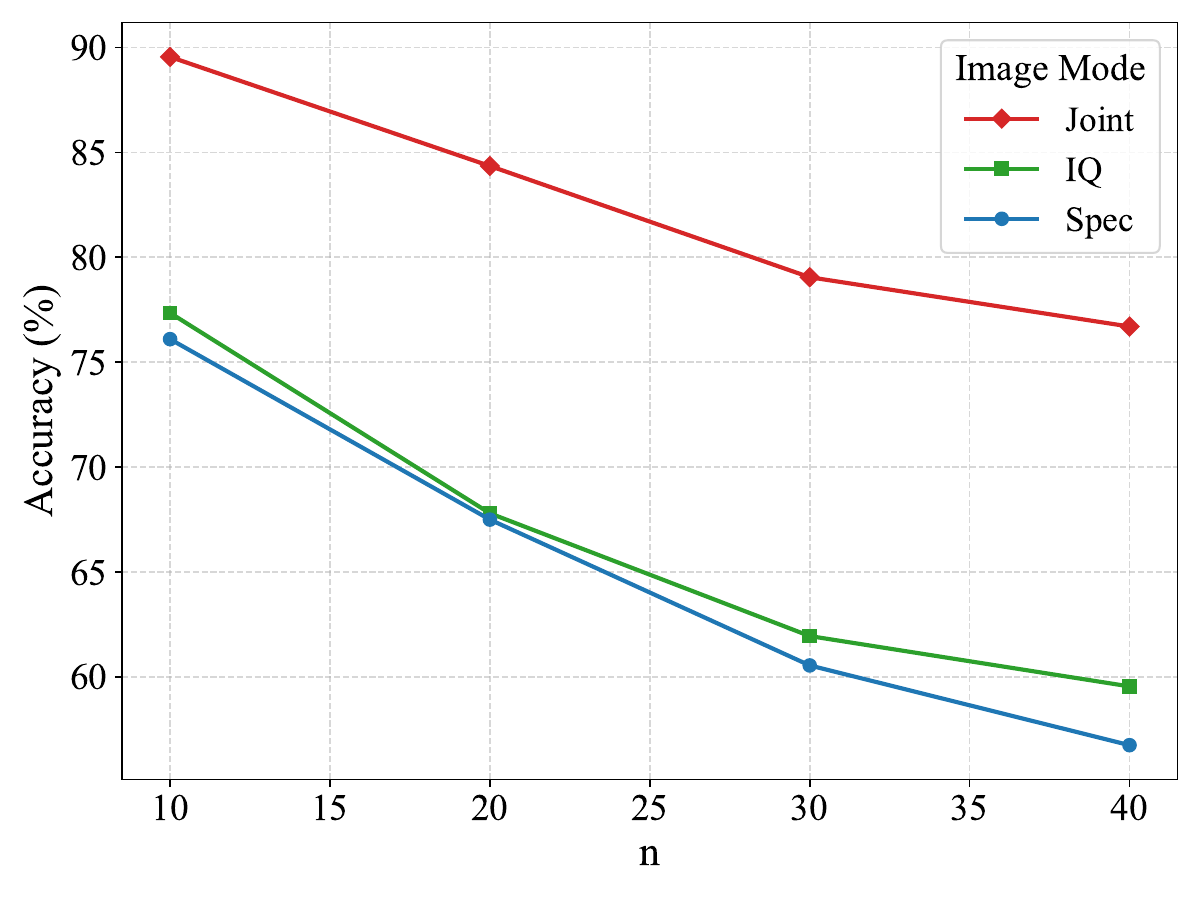}
    \caption{Accuracy vs. number of choices of Qwen2.5-VL-7B-Instruct-RF under different image modes.}
    \label{fig:accuracy_vs_num_ways}
\end{figure}

To evaluate the noise robustness of our visual representations, we fine-tuned \textcolor{black}{Qwen2.5-VL-7B-Instruct} on a mixed dataset covering signal-to-noise (SNR) values from 10 dB to 50 dB. Each VLM was separately tested at discrete SNR levels varying from 10~dB to 50~dB with a step size of 5~dB, as illustrated in Fig.~\ref{fig:SNR}. Across all image modes, we noticed a steady improvement of accuracy with higher SNR levels, confirming that each representation benefits from improved signal quality. Notably, the joint representation experiences a faster increase rate than the spectrogram-only and IQ-only modes, improving from approximately 62.20\% at 10 dB to about 86.10\% at 35 dB, after which performance plateaus. This trend reflects the fact that the IQ-only representation, which is highly sensitive to temporal fluctuations, becomes ineffective under heavy noise (e.g., at 10 dB) and thus cannot exploit the full signal structure. By contrast, the joint representation combines both spectral stability and temporal dynamics and is therefore better able to leverage the improved conditions at higher SNR values. These results demonstrate that the joint representation consistently delivers superior noise robustness.

\subsection{Out-of-Vocabulary Modulation Evaluation}

To further assess generalization under open-set conditions, we ran additional OOV experiments. We randomly excluded 4, 8, and 16 modulation classes from the full set during training while retaining the full label space at evaluation. This setup tests whether models rely on learned visual–semantic infos rather than closed-set memorization. As shown in Fig.~\ref{fig:OOV}, accuracy decreases for all image modes as the number of unseen classes grows. The joint representation maintains the highest absolute accuracy at every OOV level (e.g., from 89.55\% with no OOV classes to 78.90\% with 16 OOV classes), whereas spectrogram-only is somewhat less sensitive (from 76.10\% to 68.15\%) but remains lower in absolute performance. The IQ-only mode exhibits intermediate sensitivity and overtakes spectrogram-only as more classes are withheld.  Joint time–frequency cues yield the strongest OOV recognition overall, but mitigating their slightly higher sensitivity to open-set shift is a promising direction for future work.

\subsection{Impact of Number of Classes in VQA}

We primarily evaluate 10-way classification, but also examine how accuracy scales as the number of choices increases. Fig. \ref{fig:accuracy_vs_num_ways} shows accuracy versus the number of candidate classes for Qwen2.5-VL-7B-Instruct-RF. As expected, the accuracy of all representations decreases as the number of classes increases. As observed in other evaluations, joint representation outperforms the others largely with a gain of around \(15\%\) in average while the performance of IQ-only representation slightly dominates the spectrogram-only one. The decline is mainly due to hard negatives from the same modulation family that differ only by parameters (e.g., OFDM subcarrier count, QAM constellation size), for which the generic visual encoder might yield highly similar features. This suggests current VLM vision encoders might lack fine‑grained discrimination, motivating future works on pretraining RF-aware visual encoder on temporal and spectral views.

\subsection{Comparison with CNN/Transformer Baselines}

\begin{table}[t!]
\label{tab:scheduler_comparison}
\vspace{0.1in}
\centering
\small
\setlength{\tabcolsep}{4pt}     
\renewcommand{\arraystretch}{1.1}
\resizebox{\columnwidth}{!}{%
\begin{tabular}{lcccc}
\toprule
\makecell{Model} & Epoch 1 & Epoch 2 & Epoch 3 & Epoch 30  \\
\midrule
\makecell{Qwen2.5-VL-7B-Instruct}  & 48.40\% & 53.65\% & 56.40\%   & N/A \\
\makecell{XCiT \cite{el2021xcit}}  & 18.70\% & 28.20\% & 34.50\% & 53.8\% \\
\makecell{ResNet \cite{he2016resnet}}  & 14.10\% & 30.40\% & 33.90\% & 52.7\% \\
\bottomrule
\end{tabular}
}
\caption{Average accuracy vs. number of epochs for XCiT and ResNet and Qwen2.5-VL-7B-Instruct.}
\end{table}

Here we compare proposed VLM-based approach to CNN baselines including XCiT \cite{el2021xcit} and ResNet \cite{he2016resnet} for all 57 modulation classifications. Under identical training setting, a fine‑tuned VLM reaches 56\% accuracy after only 3 epochs, whereas a CNN baseline trained from scratch requires 30 epochs to reach 53.8\%. Beside, VLM achieves an accuracy of \textcolor{black}{48.4\%} compared to 18.7\% of XCiT after only one epoch. This suggests that, in our setting, the pretrained VLM exhibits substantially higher sample/compute efficiency, where it adapts to RF modulation patterns with an order of magnitude fewer optimization steps, benefiting from rich visual priors learned at scale. We therefore view VLMs as strong few‑epoch learners for RF spectrograms compared to task‑specific CNNs. From a system perspective, this efficiency and flexibility make VLMs a compelling backbone for RF‑aware intelligence in settings where data, compute, or task definitions evolve over time.


\subsection{Modulation Classification with Explanations}

\begin{figure}[t!]
\vspace{0.05in}
\centering
\begin{mdframed}[linecolor=black, linewidth=2pt, roundcorner=10pt]
\textbf{RF Modulation Classification with Explanation}
\hrule height 0.5pt

\vspace{0.08in}
\textbf{User:} Select the correct modulation class in \textbf{[‘qpsk’, ‘8gmsk’, ‘ook’, ‘fm’, ‘512qam-cross’, ‘4msk’, ‘4fsk’, ‘ofdm-180’, ‘8msk’, ‘16psk’]} for following RF signal with your rationale:

\begin{center}
  \includegraphics[width=0.5\linewidth]{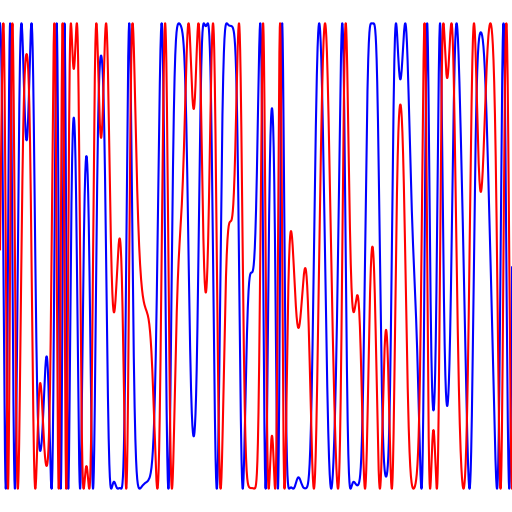}
\end{center}

\textbf{VLM:} The waveform corresponds to a modulation scheme with complex temporal patterns beyond simple harmonics. The presence of harmonics spaced at intervals corresponding to the fundamental frequency indicates a bandwidth-rich signal class. Considering the harmonics and temporal structure, the most likely class from the given list is ‘fm’.
\end{mdframed}
\caption{RF modulation classification with explanation.}
\label{fig:mcq_vqa_rationale_demo}
\end{figure}

Having established that fine-tuned VLMs can accurately classify RF modulations across different image modes, we next investigate how these models reason about their decisions. To achieve this, we prompt the models to generate  textual explanations with each prediction. As illustrated in Fig.~\ref{fig:mcq_vqa_rationale_demo}, Qwen2.5-VL-7B-Instruct-RF correctly identifies an FM signal from an IQ-segment input and provides a reasonable justification related to observing the signal structure. Notably, no explanation supervision was used during fine-tuning, yet the model often articulates domain-relevant insights.
Incorporating grounding mechanisms, such as highlighting the visual evidence, testing counterfactual examples, or verifying through retrieval, offers a promising direction to further improve accuracy and trustworthiness of the explanations.

\section{Conclusion}
\label{sec:conclusion}
In this work, we aimed to evaluate whether off‑the‑shelf VLMs can recognize RF modulations, when RF waveforms are converted into different visual representations. We demonstrated accordingly that, while vanilla VLMs show little prior ability and few‑shot in‑context learning remains ineffective, lightweight parameter‑efficient fine‑tuning within our RF‑to‑image pipeline enables them to perform competitive modulation classification without any architectural changes. Pairing a spectrogram with a short IQ trace yields the best accuracy, strong noise robustness, non‑trivial OOV generalization, and clear human-readable explanations, building a practical bridge between RF data and general-purpose VLMs. Future work will explore more realistic datasets and richer channel conditions, and extend the approach to wideband, multi-label, and streaming RF scenarios.

\bibliographystyle{IEEEtran}
\bibliography{ref} 

@article{openai20245gpt5,
  title={{Introducing GPT-5}},
  author={OpenAI},
  journal={OpenAI Blog},
  year={2025}
}

@inproceedings{brown2020gpt3,
  title        = {{Language Models are Few-Shot Learners}},
  author       = {Brown, Tom B. and others},
  booktitle    = {Advances in Neural Information Processing Systems (NeurIPS)},
  year         = {2020},
}

@article{liu2023llava,
  title={{Visual instruction tuning}},
  author={Liu, Haotian and others},
  journal = {Advances in Neural Information Processing Systems},
  volume={36},
  pages={34892--34916},
  year={2023}
}

@article{zhao2020comprehensive,
  title        = {{A Comprehensive Survey of 6G Wireless Communications}},
  author       = {Zhao, Yang and others},
  journal      = {arXiv preprint arXiv:2101.03889},
  year         = {2020},
}

@article{6gIntelligentNetwork2024,
  title        = {{6G: The Intelligent Network of Everything -- A Comprehensive Vision and Tutorial}},
  author       = {Giordani, Marco and others},
  journal      = {arXiv preprint arXiv:2407.09398},
  year         = {2024},
}

@article{wirelessllm2024,
  title        = {{WirelessLLM: Empowering Large Language Models Towards Wireless Intelligence}},
  author       = {Shao, Jiawei and others},
  journal      = {arXiv preprint arXiv:2405.17053},
  year         = {2024},
}

@article{boegner2022torchsig,
  title={{Large scale radio frequency signal classification}},
  author={Boegner, Luke and others},
  journal={arXiv preprint arXiv:2207.09918},
  year={2022}
}

@ARTICLE{zou2025telecomgpt,
  author={Zou, Hang and others},
  journal={IEEE Transactions on Machine Learning in Communications and Networking}, 
  title={{TelecomGPT: A Framework to Build Telecom-Specific Large Language Models}}, 
  year={2025},
  volume={3},
  number={},
  pages={948-975},
}

@ARTICLE{bariah2024next,
  author={Bariah, Lina and Zhao, Qiyang and Zou, Hang and Tian, Yu and Bader, Faouzi and Debbah, Merouane},
  journal={IEEE Communications Magazine}, 
  title={{Large Generative AI Models for Telecom: The Next Big Thing?}}, 
  year={2024},
  volume={62},
  number={11},
  pages={84-90},
}

@ARTICLE{hao2025CST,
  author={Zhou, Hao and others},
  journal={IEEE Communications Surveys \& Tutorials}, 
  title={{Large Language Model (LLM) for Telecommunications: A Comprehensive Survey on Principles, Key Techniques, and Opportunities}}, 
  year={2025},
  volume={27},
  number={3},
  pages={1955-2005},
}

@inproceedings{dettmers2023qlora,
 author = {Dettmers, Tim and others},
 booktitle = {Advances in Neural Information Processing Systems},
 pages = {10088--10115},
 title = {{QLoRA: Efficient Finetuning of Quantized LLMs}},
 volume = {36},
 year = {2023}
}

@article{bai2025qwen25vl,
  title={{Qwen2.5-VL Technical Report}},
  author={Bai, Shuai and others},
  journal={arXiv preprint arXiv:2502.13923},
  year={2025}
}

@article{zhu2025internvl3,
      title={{InternVL3: Exploring Advanced Training and Test-Time Recipes for Open-Source Multimodal Models}}, 
      author={Jinguo Zhu and others},
      journal={arXiv preprint arXiv:2504.10479},
      year={2025},    
}

@inproceedings{dixit2024vision,
  title={Vision Language Models Are Few-Shot Audio Spectrogram Classifiers},
  author={Dixit, Satvik and others},
  booktitle={Audio Imagination: NeurIPS 2024 Workshop AI-Driven Speech, Music, and Sound Generation},
  year={2024},
}

@article{el2021xcit,
  title={Xcit: Cross-covariance image transformers},
  author={El-Nouby and others},
  journal={arXiv preprint arXiv:2106.09681},
  year={2021}
}

@inproceedings{he2016resnet,
  title={Deep residual learning for image recognition},
  author={He, Kaiming and Zhang, Xiangyu and Ren, Shaoqing and Sun, Jian},
  booktitle={Proceedings of the IEEE conference on computer vision and pattern recognition},
  pages={770--778},
  year={2016}
}


\end{document}